\documentstyle[prb,aps,epsfig]{revtex}
\begin{document}
\draft
\title{Absence of lattice strain anomalies at the electronic topological
transition in zinc at high pressure}
\author{Gerd Steinle-Neumann\footnote{Electronic address: gerd@umich.edu} and Lars Stixrude}
\address{Department of Geological Sciences, University of Michigan, Ann Arbor, MI 48109-1063}
\author{Ronald E. Cohen}
\address{Geophysical Laboratory and Center for High Pressure Research, Washington, DC 20015-1305}
\date{submitted to Phys.\ Rev.\ B, May 15, 2000, revised July 24, 2000; accepted October 7, 2000}
\maketitle
\begin{abstract}
High pressure structural distortions of the hexagonal close packed (hcp) element
zinc have been a subject of controversy. Earlier experimental results and
theory showed a large anomaly in lattice strain with compression in zinc at
about 10 GPa which was explained theoretically by a change in
Fermi surface topology. Later hydrostatic experiments showed no such anomaly,
resulting in a discrepancy between theory and experiment. We have computed
the compression and lattice strain of hcp zinc over a wide range of
compressions using the linearized
augmented plane wave (LAPW) method paying special attention to k-point
convergence. We find that the behavior of the lattice strain
is strongly dependent on k-point sampling, and with large k-point sets the
previously computed anomaly in lattice parameters under
compression disappears, in agreement with recent experiments.
\end{abstract}

\section{Introduction}
\label{intr}

Zinc and cadmium are
unique among the hexagonal close-packed (hcp) transition metals in that the 
axial ratio ($c/a=1.856$ for zinc and 1.886 for cadmium) is far from the ideal 
value defined by hard sphere packing ($c/a=\sqrt{8/3}=1.633$).
Upon compression, the axial ratio decreases towards the ideal value.
Lynch and Drickamer \cite{lynch} first observed that the decrease in $c/a$ 
with increasing pressure was not smooth; subsequent experiments yielded
inconsistent results on the nature of this anomaly 
\cite{mcwhan,schulte1,take1,schulte2,morgan,take2}.
Takemura confirmed the anomaly using a methanol-ethanol-water mixture 
\cite{take1,take2} (MEWM) as a pressure medium in diamond anvil cell 
experiments: he observed the $a$-axis expanding
over a small range of compression, yielding a rapid decrease of the axial ratio 
$c/a$.  {\it Ab initio} computations found similar behavior 
\cite{menak,fast,nov1} and provided an explanation for the anomaly by means 
of changes in the 
Fermi surface topology under compression\cite{fast,nov2}.
Takemura recently repeated his experiments but using helium as pressure 
medium\cite{take3} which is more nearly hydrostatic than MEWM, 
but found that both axes compressed monotonically with no anomaly in $c/a$,
contrary to his earlier experiments and theory.

The most recent experimental results call previous theoretical studies 
\cite{menak,fast,nov1} into question. All previous theoretical studies 
show the anomaly in the axial ratio whether the local density 
approximation (LDA) \cite{menak,fast} or the generalized gradient approximation 
(GGA) \cite{nov1} to the exchange correlation potential is used. 
The anomaly has been connected to changes in the 
electronic structure \cite{fast,nov2}: 
Fast {\it et al.\ }\cite{fast}
observe one electronic topological transition (ETT) at the high symmetry
point K on the Brillouin zone boundary forming a ellipsoidal piece (needle) 
in the Fermi surface.
Novikov {\it et al.\ } \cite{nov2} see at least one additional ETT at 
approximately the same compression, also at K, where disconnected 
pieces form a three-leg structure of the Fermi surface along the K-M 
directions upon compression.
Depending on $c/a$ with compression Novikov {\it et al.\ }  \cite{nov2}
propose one more ETT at L (butterfly) reconciling previous contradictory 
results from first principles calculations\cite{steiner}.
All previous computations were performed with typical Brillouin zone sampling
(no more than 1000 k-points in the irreducible wedge of the Brillouin zone).

In an attempt to understand the discrepancy of the hydrostatic experiments
\cite{take3} and previous computational results \cite{fast,nov1} we calculate 
the equation of state, lattice constants, and electronic structure of zinc 
over a wide compression range from  first principles paying particular 
attention to the convergence of the calculations with respect to reciprocal 
space integration.  In section \ref{meth} we introduce the method used and 
elaborate the computational details of our first principles calculations.
Section \ref{res} focuses on our results for 
the equation of state, lattice constants, and electronic structure. 
We compare our results to experiments at ambient conditions and high pressure 
and to previous theoretical work. Discussion and conclusions follow.

\section{Method}
\label{meth}

We investigate the energetics of hcp zinc using the
full-potential linearized-augmented plane-wave method (LAPW) \cite{singhbook}
with GGA\cite{PBE}. Core states are treated
self-consistently using the full Dirac equation for the spherical part of the
potential, while valence states are treated in a semirelativistic approximation
neglecting spin-orbit coupling.  $3s$, $3p$, $3d$, $4s$, and $4p$ states are 
treated as valence electrons. The muffin-tin radius $R_{MT}$ 
is 2.0 Bohr over the whole compression range considered.

We perform calculations at three sets of Brillouin zone sampling, 24x24x24,
32x32x32, and 48x48x48 special k-points \cite{monk}, yielding 732,
1632, and 5208 k-points in the irreducible wedge of the Brillouin zone for 
the hcp lattice, respectively. The lowest k-point sampling is comparable to 
the previous GGA study \cite{nov1} while the latter two are much denser than
any previously published results. The size of the basis is set by
R$_{MT}$K$_{max}=9.0$, where K$_{max}$ is the largest reciprocal space vector. 
We use Fermi broadening with an electronic temperature of 5 mRy. 
For the densest k-point mesh we also perform calculations without 
electronic broadening for a selected subset of volumes and do not see any 
significant change in our results: equilibrium axial ratios remain within 
$\pm$ 0.005, the uncertainty of our results.

We carry out total energy calculations over a wide range of unit cell volumes.
At each volume we perform calculations for several different values of the
axial ratio and find the equilibrium structure by fitting the results to
a quadratic.
The equation of state is obtained by describing the 
energy-volume curve with a third order expansion in Eulerian finite 
strain\cite{birch}.
We consider unit cell volumes ranging 
from 60-110 Bohr$^3$ for 24x24x24 k-points and focus on the range in which the
anomaly in $c/a$ occurs (90-102.5 Bohr$^3$) for 32x32x32 and 48x48x48
k-point meshes. 

\section{results}
\label{res}

\subsection*{Equation of State}

A comparison of the pressure-volume relation between our results and
static experiments \cite{take2,take3} (Fig.\
\ref{eos} and Table \ref{EOS_table}) shows good agreement at low pressure.
At higher pressure theory differs significantly from the results of the 
MEWM diamond anvil cell experiments \cite{take2};
this is consistent with previous theoretical results \cite{nov1}.

To investigate whether non-hydrostaticity may be responsible for the discrepancy
we also compare to the results of shock wave experiments \cite{marsh} where
hydrostaticity is readily achieved \cite{brown}
(Fig.\ \ref{eos} and Table \ref{EOS_table}). The Hugoniot is reduced to a 0 K 
isotherm by solving the Rankin-Hugoniot equation \cite{zharkov}. 
We estimate the thermal pressure ($P_{th}=\gamma E_{th}/V$)
along the Hugoniot, with $\gamma$ the Gr\"uneisen parameter and 
$E_{th}$ the thermal energy .
We approximate the thermal energy by the Dulong Petit law ($C_{V}^{lat}=3R$); 
the electronic contribution to the thermal pressure is negligible 
(the temperature along the Hugoniot is less than 2000 K).
We assume the Gr\"uneisen parameter is proportional to compression 
($\gamma=\gamma_0 V/V_0$) with $\gamma_0$ its zero pressure value
evaluated from the thermodynamic definition 
($\gamma=\alpha K_T/C_V \rho$), where the thermal expansivity $\alpha$, the
isothermal bulk modulus $K_T$, specific heat $C_V$ and density $\rho$ at 
zero pressure are taken from the literature \cite{james}.

The reduced Hugoniot agrees with our GGA results much better than the static 
experiments; differences in volume are less than 1.5\%. 
The large discrepancy between the static and shock wave experiments 
indicates that the MEWM experiments \cite{take2} may be biased by 
non-hydrostaticity.

\subsection*{Lattice Constants}

Total energy as a function of axial ratio for the 24x24x24 k-point mesh shows 
an unusually large scatter about the quadratic fit in $c/a$ (Fig.\ \ref{975}). 
With increasing number of k-points the scatter decreases and the minimum 
becomes better defined. In contrast to the previous GGA results \cite{nov1} 
we do not see multiple minima in $c/a$ for any volume and find the axial
ratio reliably resolved to within $\pm 0.005$, within the symbol size in Fig.\ 
\ref{eos}. The curvature of energy as a function of $c/a$ varies 
considerably for the different k-point meshes, showing that elastic constants
will also be strongly dependent on k-point sampling, as the shear elastic
constant ($C_S$) is related to this strain \cite{steinle}.

The development of the axial ratio $c/a$ with compression differs for the three
sets of computations considerably (Fig.\ \ref{eos}). For 24x24x24 k-points
we see an anomaly similar to that in the MEWM experiments\cite{take1,take2}: 
after an initial
linear decrease in the axial ratio (102.5-95 Bohr$^3$) the slope in $c/a$ 
steepens (95-90 Bohr$^3$) before decreasing again at higher pressures. 
The dependence of $c/a$ on compression for k-point meshes of 32x32x32 and 
48x48x48 is much smoother; the anomaly in $c/a$ has disappeared. 
The difference between experiment and theory
is less than 4\% in $c/a$ which is typical of all electron calculations 
\cite{fast95}.
At higher compression (V$<$ 70 Bohr$^3$) the theoretical value 
smoothly approaches 1.61; the MEWM experiments converge to 1.59.

The nature of the anomaly is revealed by considering the 
lattice constants separately (Fig.\ \ref{ca}). The $c$-axis compresses 
monotonically with decreasing volume in all computations and experiments 
considered. Theory overpredicts $c$ by less than 2\%, and there is little 
difference in $c$ for the two denser k-point meshes. 
An expansion of the $a$-axis for the 24x24x24 k-point calculations and the
MEWM experiments cause the anomaly in $c/a$ (Fig.\ \ref{ca}). For the 
two denser k-point meshes $a$ compresses monotonically; for volumes smaller
than $V=95$ Bohr$^3$ $a$ follows a linear trend with the same slope as the 
helium experiments. 
For volumes greater than V=95 Bohr$^3$, $a$ is less compressible than it is at
higher pressure.
For the the larger k-point samplings the calculations 
underestimate $a$ by less than 1\%, while with 24x24x24 k-point the maximum
difference is approximately 1.5\%. 

To illustrate this point further we evaluate the linear compressibility
for the two axes $k_x=-(1/x)(\partial x/\partial P)$ (with $x=a,c$) 
for our results and the static experiments \cite{take2,take3} 
using central differences (Fig.\ \ref{ka}). 
Our results for 24x24x24 k-points show an anomaly in $k_a$ similar in 
character and magnitude to that found in the MEWM experiments.
For denser k-point sampling the anomaly in $k_a$ is shifted towards lower 
pressure and its magnitude decreases with increasing number of k-points. 
For $k_c$ an anomaly exists as well for both the MEWM experiments \cite{take2}
and the calculations with the smallest k-point mesh, it is however less 
pronounced than for $k_a$ and is absent from the results for
the two denser k-point meshes.

\subsection*{Electronic Structure}

The band structure of zinc under compression changes considerably 
from $V=102.5$ Bohr$^3$ to $V=95$ Bohr$^3$ (Fig.\ \ref{bands}).
The electronic structure is 
in excellent agreement with the previous GGA results \cite{nov1}.
The major change in band structure occurs at the high symmetry point K on the 
Brillouin zone boundary where three bands (K$_7$,K$_8$,K$_9$\cite{stark})
cross the Fermi energy under compression, changing the topology of the 
Fermi surface. From the band structure we see the needle around K
and also the connection of the three-leg piece at K. 
Focusing on the development of the band structure at K we consider the 
eigenvalues of the K$_7$, K$_8$, and K$_9$ states (Fig.\ \ref{dE}). 
For the 24x24x24 k-point calculations these bands show a quadratic volume
dependence and cross the Fermi energy at V=97.5 Bohr$^3$ (K$_7$ and K$_8$) 
and V=97 Bohr$^3$ (K$_9$).
For the two denser k-point meshes the eigenvalues depend 
linearly on volume and the crossing points are indistinguishable for 32x32x32 
and 48x48x48 k-points. The crossings occur at slightly higher volume than
for the 24x24x24 k-point mesh (V=98 and 97.5 Bohr$^3$ for K$_7$, K$_8$, 
and K$_9$, respectively), but the difference is small compared 
with the effect of k-point sampling on the lattice parameters.

\section{Discussion}
\label{dis}

The ETT discussed in the last section has previously been invoked as an
explanation for the anomaly in $a$-axis compressibilities \cite{fast,nov2}.
In contrast to these studies we find that the occurrence of the ETT is 
independent of the calculated anomaly in $a$-axis compressibility,
as the location of abnormal compression of $a$ shifts with increasing 
k-point sampling towards higher volumes (Figs.\ \ref{ca} and \ref{ka}) 
while the ETT always occurs at approximately the same volume (Fig.\ \ref{dE}).
The anomaly in $a$-axis compression seen in previous calculations appears to 
be a consequence of insufficient k-point sampling. The results presented here
for $a$ (Fig.\ \ref{ca}) and linear compressibility $k_a$ (Fig.\ \ref{ka})
suggest that even for the densest k-point mesh we use (48x48x48) the lattice 
parameters are not converged.

The discrepancy between the MEWM and helium experiments can be attributed
to freezing of the MEWM pressure medium which is known to occur at about 
10 GPa \cite{pier}. Freezing substantially increases the 
non-hydrostatic component of stress as recognized previously
in high pressure experiments on forsterite (Mg$_2$SiO$_4$)
\cite{downs}.
At room temperature helium also freezes within the pressure range of
the experiment (11.5 GPa) \cite{besson} but remains soft enough to
maintain hydrostaticity \cite{downs}. 
Recent neutron inelastic-scattering experiments under compression
\cite{klotz,overhauser} show no softening or anomaly in the phonon frequency,
supporting the monotonic compression of both axes as seen in our dense k-point
calculations and the helium experiments.

In retrospect it occurs as a fortuitous (or unfortunate) coincidence that
for {\it typical} computational parameters comparable behavior in linear 
compressibilities is found in first principles electronic
structure calculations and for experiments with non-hydrostatic conditions,
despite the fundamentally different underlying physical problem.
Following the same notion the observation of anomalies in the axial ratio 
under compression for cadmium 
in experiments \cite{take2} and theory using both LDA\cite{godwal} 
and GGA \cite{nov1} may also be an artifact due to non-hydrostatic 
conditions in the experiments and insufficient convergence with 
respect to computational parameters as well. 

The ETT, however, might have important effects on higher order physical 
properties such as elasticity. For tantalum a similar change in electronic
structure as for zinc has been found under compression which has little effect 
on the equation of state \cite{cohen} but appears in the elastic constants 
\cite{gulseren}.

\section{Conclusions}
\label{conc}

Using the first-principles LAPW method with GGA 
we calculate the equation of state, structural parameters, and 
electronic structure of zinc over a wide compression range. We perform 
calculations for three different k-point samplings of the first Brillouin zone 
(24x24x24, 32x32x32, and 48x48x48 k-points) and 
find lattice parameters, in particular the $a$-axis, strongly dependent on 
the number of k-points, while
little or no effect can be seen on the equation of state and band structure.
For lattice constants we find that a previously observed
anomaly in $a$-axis compressibility shifts to lower pressure and decreases in 
amplitude as we increase k-point sampling from 24x24x24 to 48x48x48.
This anomaly is not coupled to
a change in electronic band structure as has been proposed before; we 
observe the ETT occurring at approximately the same volume for all sets of 
computational parameters. 

The disappearance of the anomaly in lattice constants for our results is 
in agreement with recent static experiments using helium 
as a pressure medium. The remaining anomaly in $a$-axis compressibility
indicates that structural 
parameters are not fully converged even for the prohibitively large k-point
sampling we perform.

\section*{Acknowledgement}

We greatly appreciate helpful discussion with W.\ Holzapfel and K.\ Takemura 
at different stages of this project and thank H.\ Krakauer 
and D.\ Singh for use of their LAPW code. This work was supported by the 
National Science Foundation under grants EAR-9614790 and EAR-9980553 (LS), 
and in part by DOE 
ASCI/ASAP subcontract B341492 to Caltech DOE W-7405-ENG-48 (REC). Computations 
were performed on the SGI Origin 2000 at the Department of Geological Sciences 
at the University of Michigan and the Cray SV1 at the Geophysical Laboratory, 
support by NSF grant EAR-9975753 and by the W.\ M.\ Keck Foundation.


%
%
%
\epsfig{file=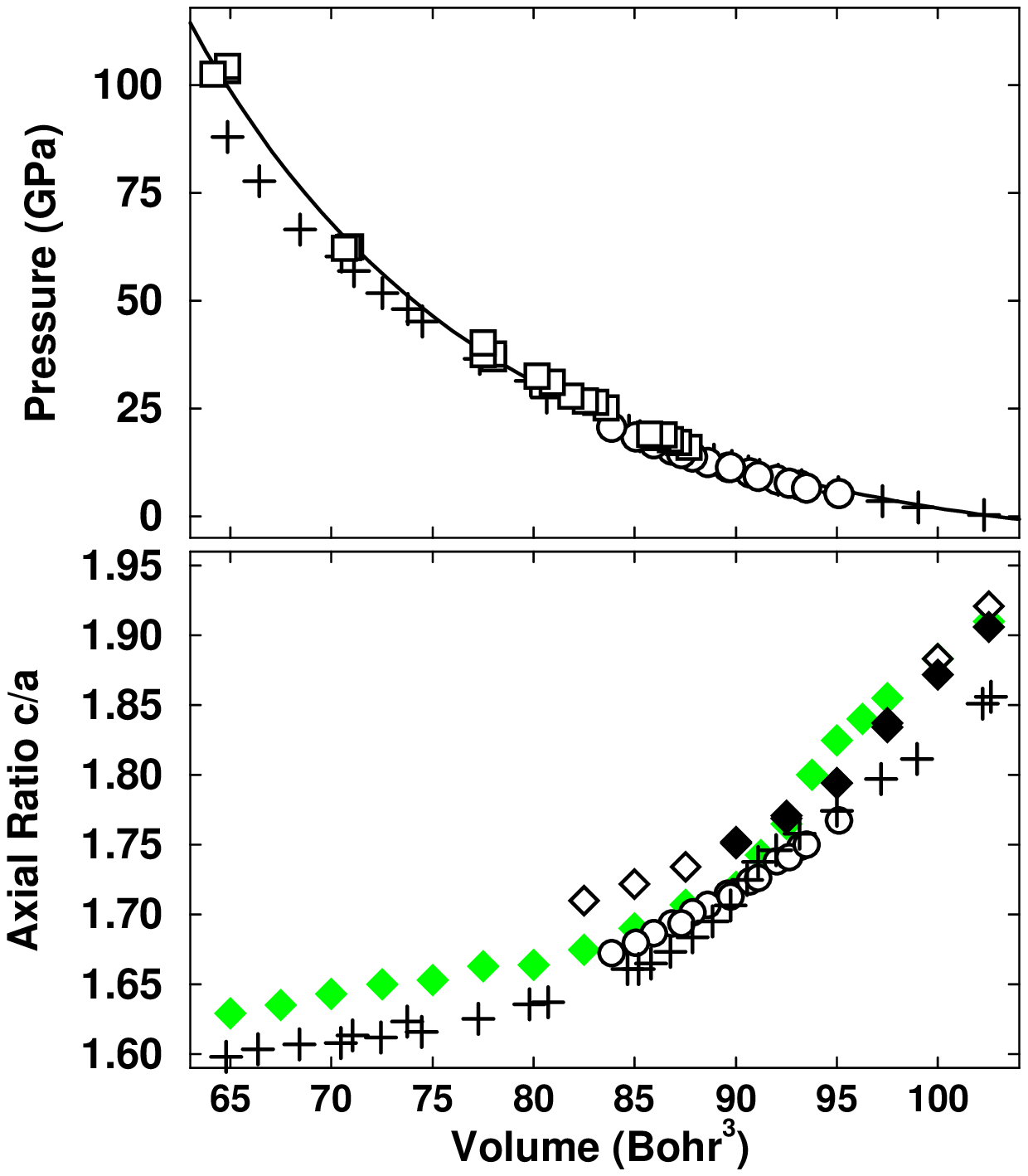,width=4in}
\begin{figure}
\caption{
Axial ratio $c/a$ and equation of state for zinc from our calculations and 
experiment. The lower panel compares our results for $c/a$ (gray diamonds 
24x24x24 k-points, open diamonds 32x32x32 k-points, and filled diamonds
48x48x48 k-points) to static experiments using methanol ethanol water  
mixture (Ref.\ \protect\onlinecite{take2}, pluses) and helium
(Ref.\ \protect\onlinecite{take3}, circles).
The equation of state for zinc is shown in the upper panel 
for our calculations (line) 
and the same two set of diamond anvil cell experiments (same symbols as above)
The open squares
show shock wave experiments (Ref.\ \protect\onlinecite{marsh}) reduced to a
0 K isotherm.
}
\label{eos}
\end{figure}
\epsfig{file=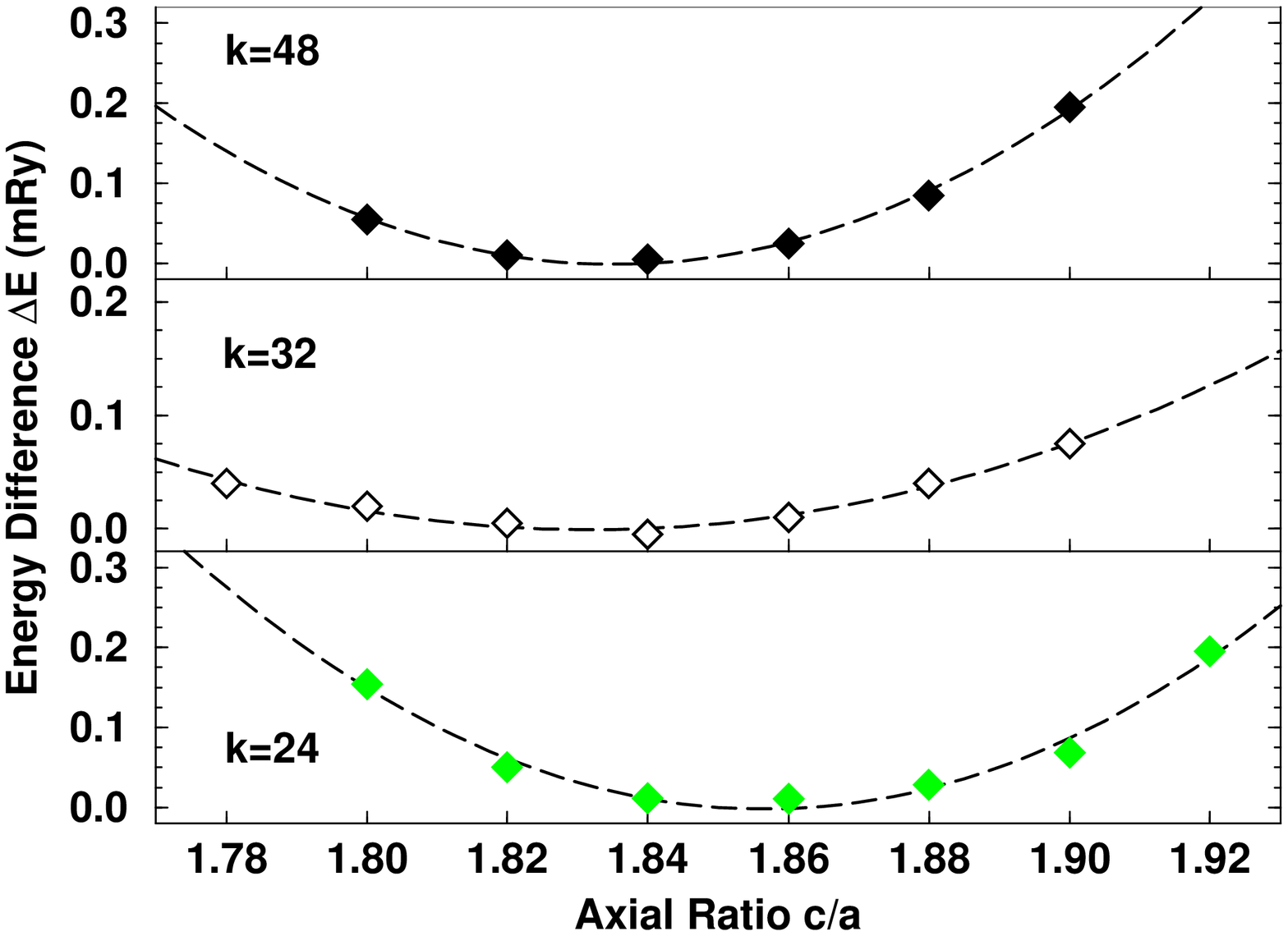,width=4in}
\begin{figure}
\caption{
Relative energies as a function of the axial ratio $c/a$ for $V=97.5$ Bohr$^3$.
The lower, middle, and upper panel shows results for 24x24x24, 32x32x32, and 
48x48x48 k-points, respectively. The lines show quadratic fits in $c/a$ to 
the results.
}
\label{975}
\end{figure}
\epsfig{file=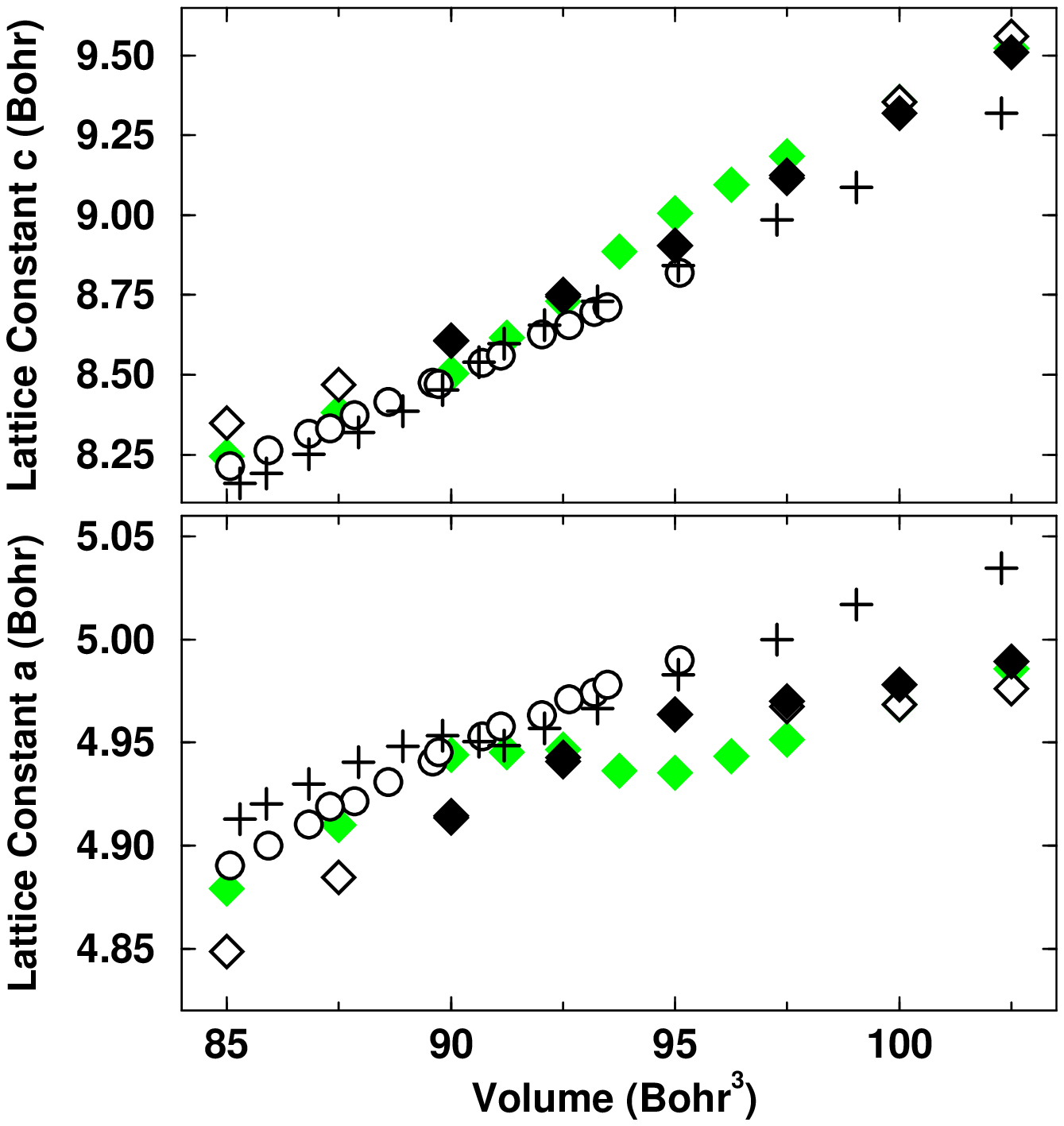,width=4in}
\begin{figure}
\caption{Compression of the two axes $a$ and $c$ in the
hexagonal cell for zinc over the compression range $V/V_0=0.80-1.0$. 
In the upper panel we compare our results (diamonds) for $c$ 
(gray 24x24x24, open 32x32x32, and filled 48x48x48 k-points) 
with the static experiments by Takemura using methanol-ethanol-water mixture
(pluses, Ref.\ \protect\onlinecite{take2}) and 
helium (open circles, Ref.\ \protect\onlinecite{take3}) as a pressure medium.
For $a$ in the lower panel the same symbols are used.  Note the approximately 
fivefold difference in range of axes compressibilities for $c$ and $a$.
}
\label{ca}
\end{figure}
\epsfig{file=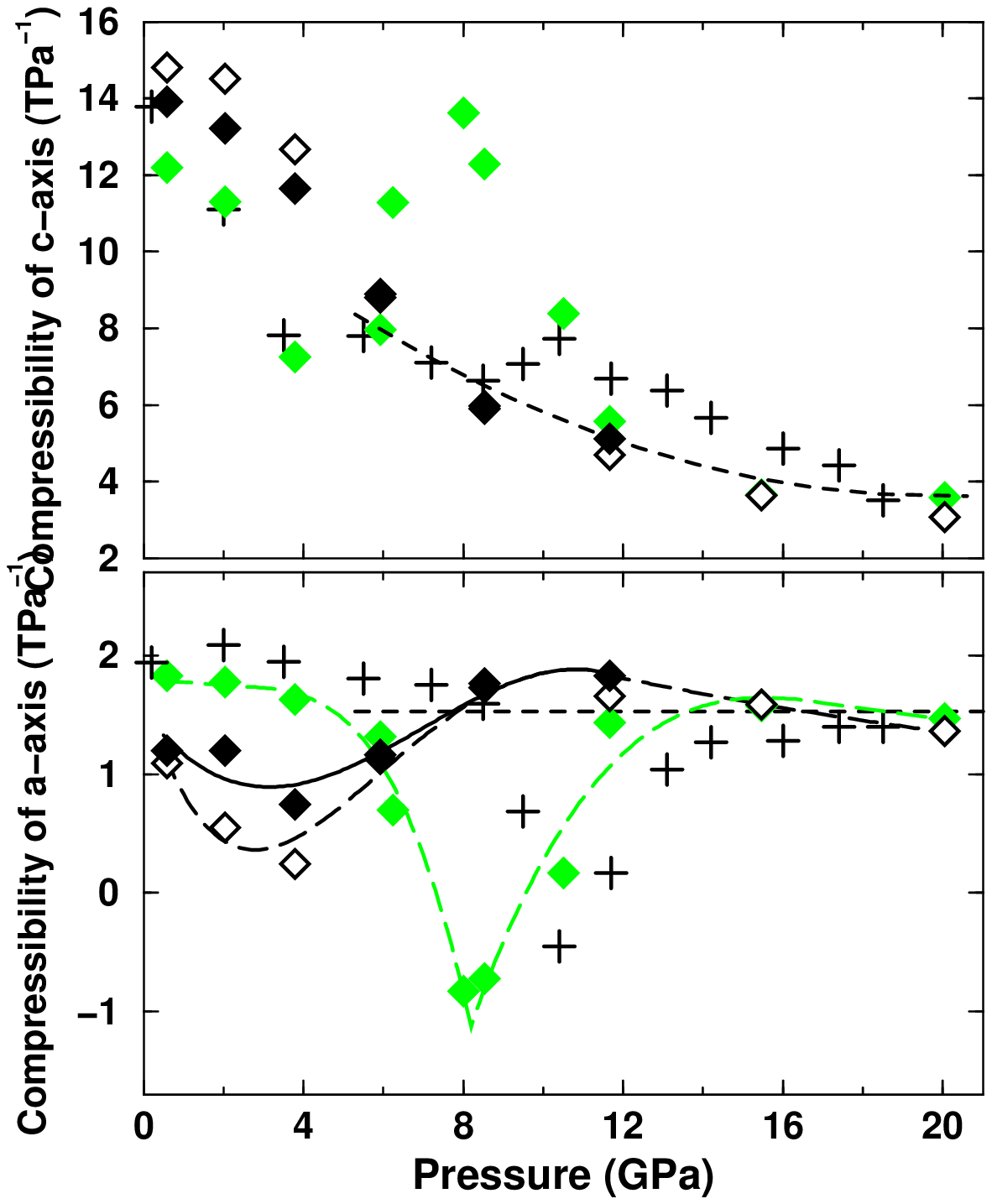,width=4in}
\begin{figure}
\caption{Compressibility of the two axes $k_a$ and $k_c$ for our results 
(k-point sampling 24x24x24 with grey diamonds, for 32x32x32
and 48x48x48 we use a fit to the results in the long dashed and solid line, 
respectively) and static 
experiments using a methanol-ethanol-water mixture (pluses, Ref.\
\protect\onlinecite{take2}) and helium (average in dashed line, Ref.\ 
\protect\onlinecite{take3}) as a pressure pressure medium.
}
\label{ka}
\end{figure}
\epsfig{file=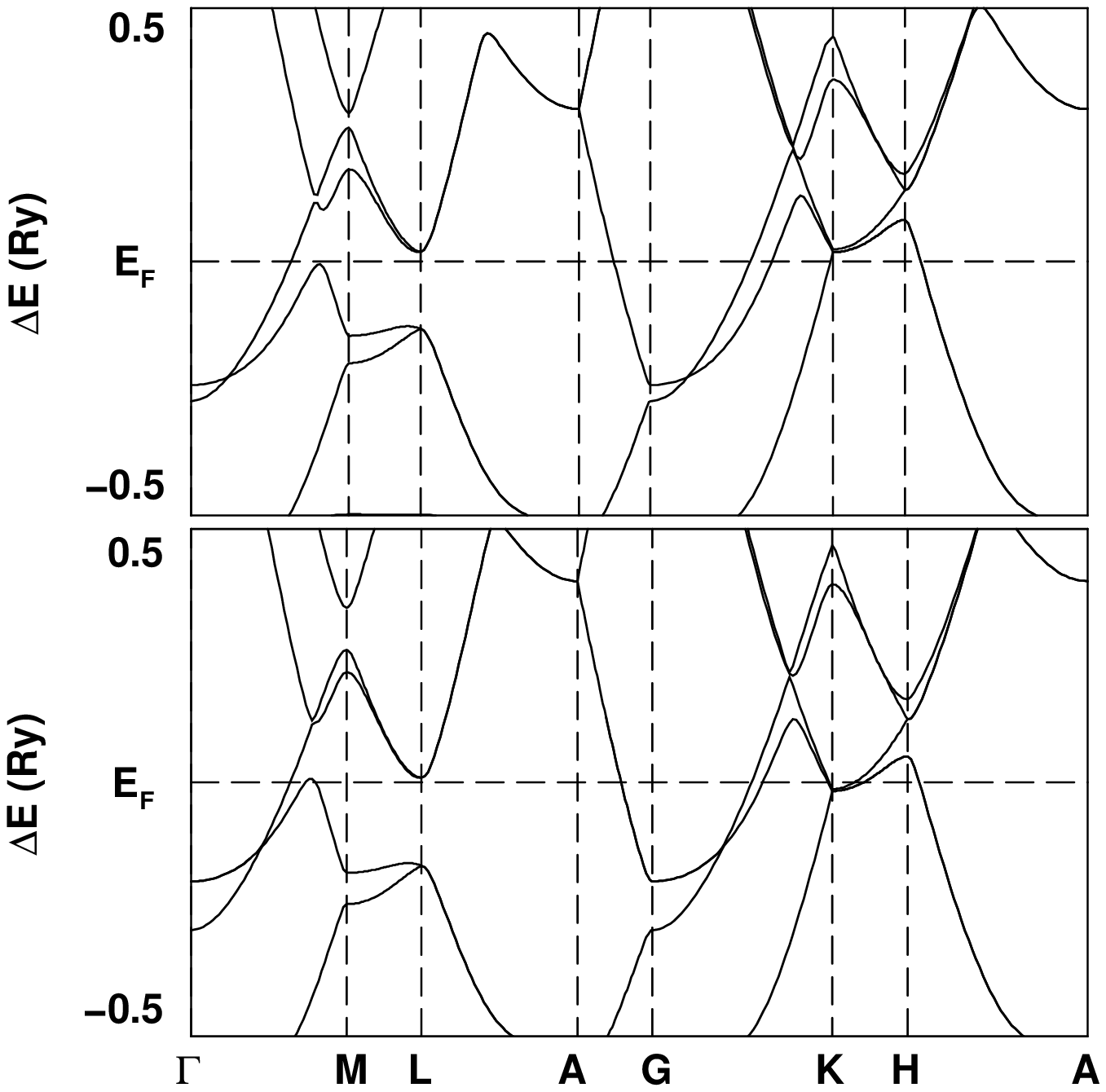,width=4in}
\begin{figure}
\caption{Band structure for zinc along high symmetry directions in the 
first Brillouin zone around the Fermi energy. The upper panel shows the band 
structure at zero pressure ($V=102.5$ Bohr$^3$), the lower panel at $V=95$ 
Bohr$^3$ at their equilibrium $c/a$ (1.91 and 1.79, respectively). 
}
\label{bands}
\end{figure}
\epsfig{file=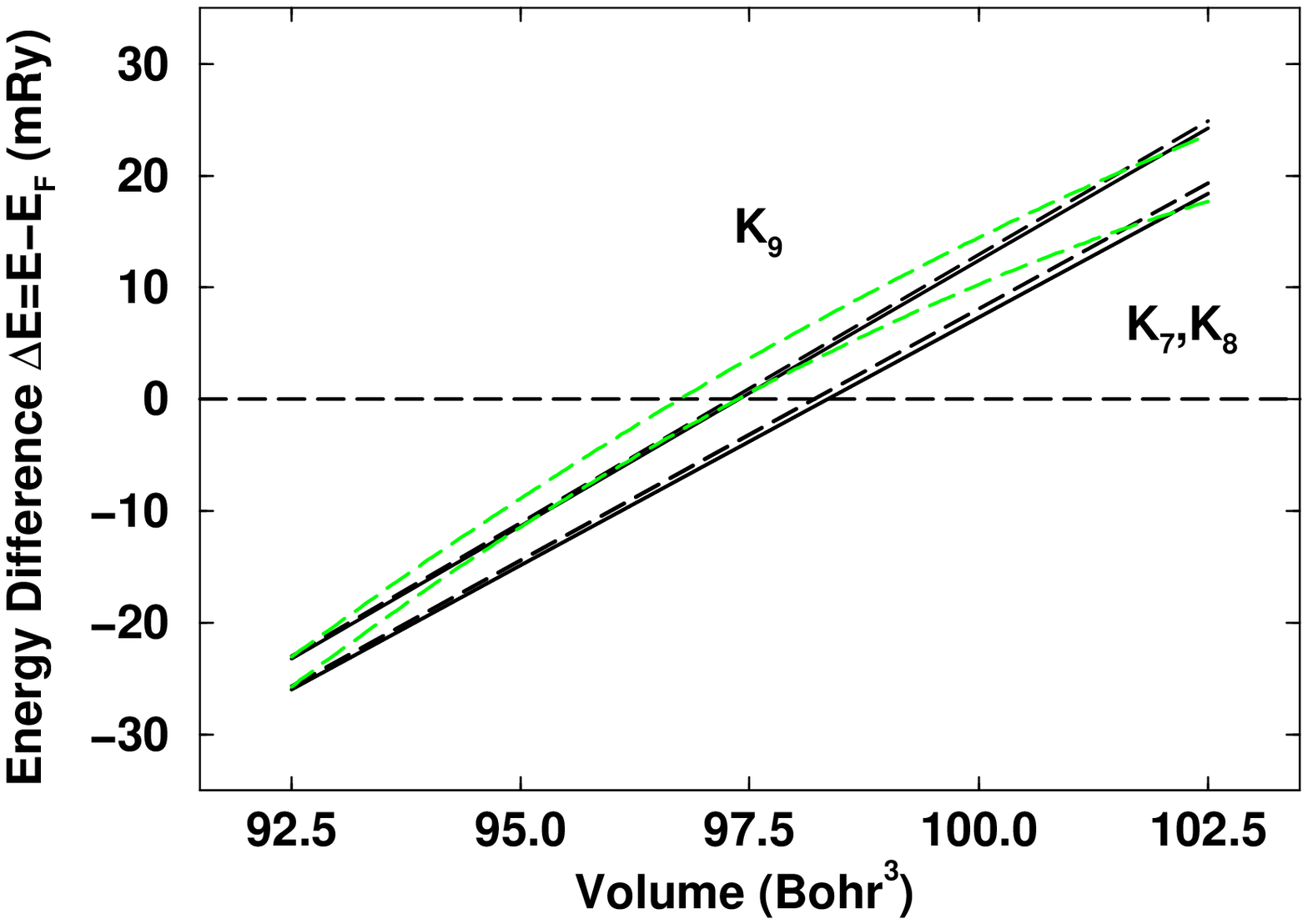,width=4in}
\begin{figure}
\caption{Energy differences of the bands to the Fermi energy at the high 
symmetry point K on the first Brillouin zone boundary as a function of 
unit cell volume. Gray dashed,
dashed, and solid lines are results for k-point sampling of 24x24x24, 
32x32x32, and 48x48x48, respectively. 
}
\label{dE}
\end{figure}
%
%
%
\newpage
\begin{table}
\caption{Equation of state parameters for our calculations (GGA) and experiment:
$V_0$, $K_0$, and $K_0^{'}$ are the equilibrium volume, bulk modulus, and its
pressure derivative at $V_0$, respectively. Due to the restricted compression 
range of the calculations with higher k-point sampling (32x32x32 and 48x48x48)
and for the static experiments with helium as a 
pressure medium, we constrain $K_{0}'$ and $V_0$.
}
\label{EOS_table} 
\begin{tabular}{lllll}
method                        & $V_0$      & $K_0$ & $K_0^{'}$ \\
                              & [Bohr$^3$] & [GPa] &           \\
\hline
GGA k=24                      & 102.8      & 64    & 5.2 \\
GGA k=32                      &            & 64    &     \\
GGA k=48                      &            & 63    &     \\
\hline
equilibrium properties       &102.6\cite{james} & 60\cite{land} & \\
reduced Hugoniot \cite{marsh}&             & 69    & 4.9 \\
experiment MEWM\cite{take2}  &             & 65    & 4.7 \\ 
experiment He\cite{take3}    &             & 61    &     \\
experiment N\cite{schulte2}  &             & 63    & 5.2 \\
\hline
FP-LMTO (GGA)\cite{nov1}     & 101.5       & 60    &    
\end{tabular}
\end{table}
\end{document}